\documentclass[amsmath,amssymb,aps,pra,twocolumn]{revtex4-1}

\usepackage{graphicx}
\usepackage{dcolumn}
\usepackage{bm}

\newcommand{\tra}{\mathrm{Tr}}
\newcommand{\be}{\begin{equation}}
\newcommand{\ee}{\end{equation}}
\newcommand{\ben}{\begin{eqnarray}}
\newcommand{\een}{\end{eqnarray}}

\begin{document}
\title{High-dimensional Angular Two-Photon Interference and Angular Qudit States}

\author{Graciana Puentes$^{1,2}$}
    \email[Correspondence email address: ]{gpuentes@df.uba.ar}
    \affiliation{1-Departamento de Fsica, Facultad de Ciencias Exactas y Naturales, Universidad de Buenos Aires, Ciudad Universitaria, 1428 Buenos Aires, Argentina,
    2-CONICET-Universidad de Buenos Aires, Instituto de Fsica de Buenos Aires (IFIBA), Ciudad Universitaria, 1428
Buenos Aires, Argentina.}

\date{\today} 

\begin{abstract}
Using angular position-orbital angular momentum entangled photons, we propose an experiment to generate maximally entangled states of
$D$-dimensional quantum
systems, the so called qudits, by exploiting correlations of parametric down-converted photons. Angular diffraction masks containing
$N$-slits in the arms of each twin photon define a qudit space of dimension $N^2$, spanned by the alternative pathways of the photons. Due to phase-matching conditions, the twin photons will pass only by symmetrically opposite angular slits, generating
maximally entangled states between these different paths, which can be detected by high-order two-photon interference fringes via coincidence counts. Numerical results for $N$ angular slits with
$N = 2, 4, 5, 6, 10$ 
are reported, corresponding to qudit Hilbert spaces of dimension $D=N^2=4,16,25, 36,100$, respectively. We discuss relevant experimental parameters for an experimental implementation of the proposed scheme using Spatial Light Modulators (SLMs), and twin-photons produced by Spontaneouos Parametric Down Conversion (SPDC). The entanglement of the qudit state can be 
quantified in terms of the Concurrence, which can be expressed in terms of the visibility of the interference fringes, or by using Entanglement Witnesses. These results provide an additional means for preparing entangled
quantum states in  high-dimensions, a fundamental resource  for quantum simulation and  quantum information protocols.
\end{abstract}


\maketitle

\section{Introduction}

\begin{figure}[t]
\centering
\includegraphics[width=0.5\textwidth]{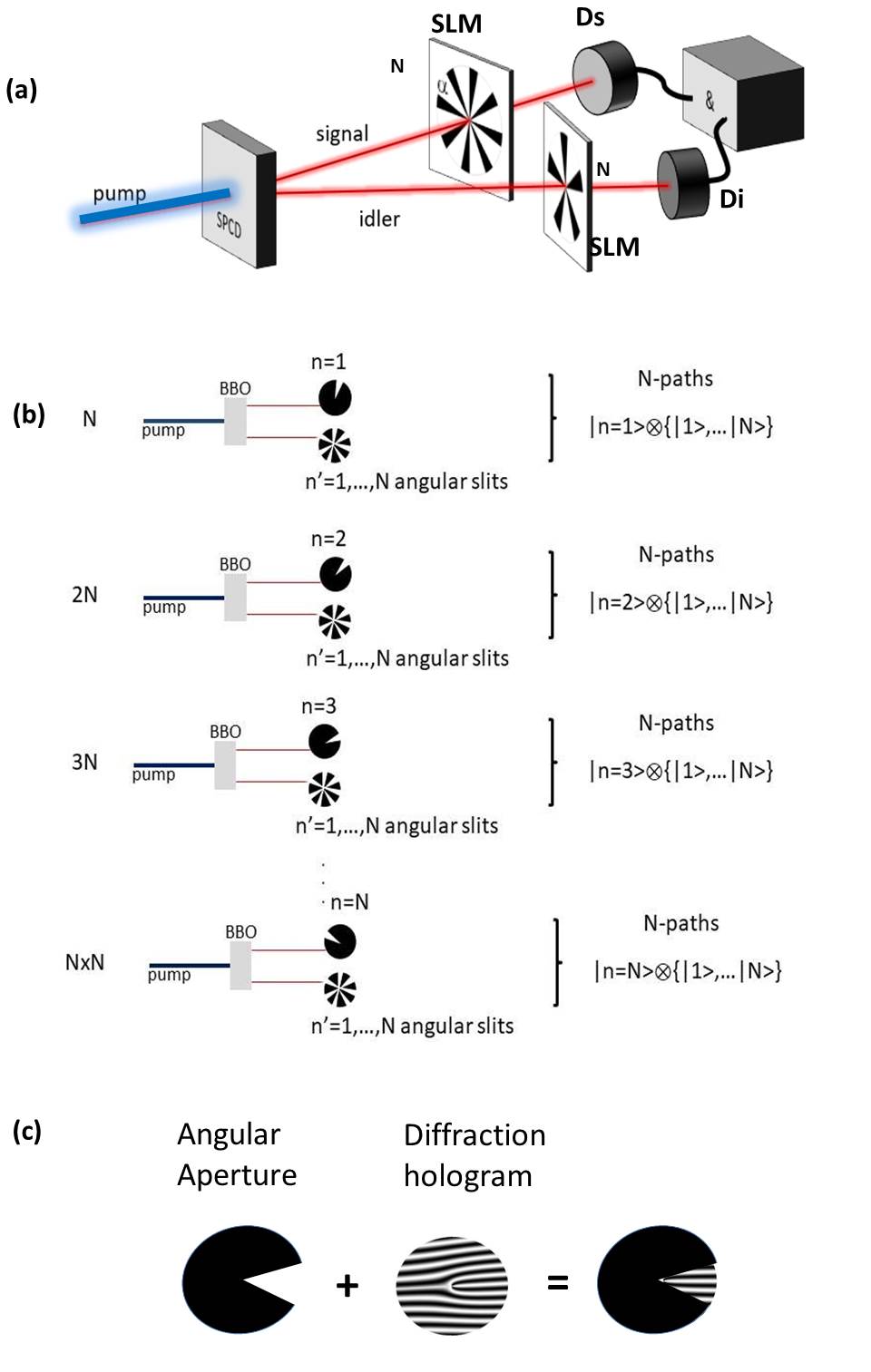}
\caption{(a) Schematic of proposed experimental setup (see text for details). (b) Two-photon multiple-path diagram showing  $N^2$ alternative paths using angular masks containing $N$ slits of width $\alpha$ and separation $\beta$, with $N(\alpha + \beta) \leq 2 \pi$. (c) Angular aperture and diffraction hologram used to analyze the OAM spectrum. Both the angular aperture and the diffraction hologram are programmed using standard Spatial Light Modulators \cite{18}.}
\label{fig:awesome_image}
\end{figure}

Twin photons produced by the non-linear process of spontaneous parametric down-conversion (SPDC) are entangled in several independent
degrees of freedom including  position and
momentum, polarization, time and energy, or angular position and orbital angular momentum (OAM). Entanglement of the two photons in a given domain gives rise to two-photon coherence, which manifests itself
as an interference effect in that particular domain.
Two-photon interference has been observed in the temporal \cite{1,2,3,4,5} and spatial \cite{6,7,8} domains.
These effects are used to test the foundations of
quantum mechanics \cite{9,10,11} and are central to many quantum information applications  \cite{12,13,14}. \\

Fourier relationship linking angular position and OAM  leads to  angular interference in the OAM-mode distribution of a photon when it passes through an angular aperture, resulting in two-photon interference in the angular domain \cite{15, 16, 17, 18, 19,20,21}. In this article, we
study high-dimensional angular two-photon interference in a scheme in
which entangled photons produced by SPDC are made to pass through multiple angular apertures in the form of $N$ angular slits, which results in path entanglement in a space of dimension $D=N^2$, the so called qudits. Using this
scheme, it is possible to demonstrate an entangled qudit state that
is based on the angular-position correlations of down-converted photons. Entangled two-qubit states are the necessary ingredients for many quantum information protocols \cite{12,13,14}, and they have previously been realized by
exploring the correlations of entangled photons in variables
including polarization \cite{22}, time bin \cite{4,5}, frequency \cite{23},
position \cite{7,8}, transverse momentum \cite{24,25}, and OAM
\cite{19,20,21}. To date, angular-position correlations of twin photons have only been demonstrated for $N=2$ angular slits \cite{56}, which represent a two-qubit system. The results presented
here extended this notion to an arbitrary number of angular slits $N$, which not only demonstrates two-photon coherence effects in
the angular domain but also provide an additional means
for preparing entangled quantum states in a high dimensional space (qudits), which is a fundamental resource for quantum information protocols.\\
 
The article is organized as follows, in Section I, we present an introduction to the problem and the analytical tools for characterizing high-dimensional interference effects in the angular position-OAM domain. Second, in Section II, we present a graphical representation of typical density matrices of pure maximally entangled states for $N=2,4,5,10$, corresponding to qudits systems of dimension $D=N^2=4,16,25,100$, and we numerically reproduce the results obtained by Kumar \emph{et al.}, PRL 2010 \cite{56} for the case $N=2$. In Section III, we present numerical results for multipath interference effects for $N=6$ angular slits, corresponding to qudits systems of dimension $D=N^2=36$, and for the case of an asymmetric configuration of angular slits $N=6$ and $M=3$, resulting in mixed states in a Hilbert space of dimension $D=18$ . Next, in Section IV we briefly introduce a scheme based on Entanglement Witnesses to estimate a lower bound on the entanglement content of angular qudits, using the Logarithmic Negativity as a measure of entanglement. In Section V, we discuss the requirements for an experimental implementation of this proposal, and we provide an estimation of the largest qudit space that can in principle be implemented with this approach using state-of-the-art Spatial Light Modulators. Finally, in Section VI we present our conclusions.\\

Let us consider the experimental setup described in Fig. 1(a). In the simplest scenario, a Gaussian pump beam produces signal ($s$) and idler ($i$) photons, by type-I degenerate spontaneous parametric down conversion (SPDC) with non-collinear phase-matching. For a pump  beam with zero OAM ($l=0$), phase matching conditions imply that the two-photon down-converted state $|\psi_{sl} \rangle$ is given by [56]:

\begin{equation}
|\psi_{sl} \rangle= \sum_{l=-\infty}^{+\infty} c_{l}|l\rangle_{s}|-l\rangle_{i},
\end{equation}

where $s$ and $i$ stand for signal and idler photons, respectively, and $|l\rangle$ represents an OAM eigen-mode of order $l$, associated with and azymuthal phase of the form  $e^{il\phi}$. $|c_{l}|^2$ represent the probability of generating signal and idler photon in the OAM mode of order $l$, where normalization condition imposes $\sum_{l=-\infty}^{+\infty}|c_{l}|^2=1$. The width of this mode probability
distribution is referred to as the spiral bandwidth of the
two-photon field [56]. Signal and idler photons are
made to pass through $N$ angular slits, as shown in
Fig. 1(a), placed in the image planes of the crystal. The
amplitude transmission functions $A_{j,n}$ of the individual angular
slits are given by: 

\begin{equation}
A_{j,n}(\phi_{j})= 1 \hspace{0.2cm} \mathrm{if}  \hspace{0.2cm} n\beta -\alpha/2 \leq \phi_{j} \leq n \beta + \alpha/2  \hspace{0.2cm} \mathrm{else}  \hspace{0.2cm} 0,
\end{equation}

where $j=(s,i)$ is the label for signal and idler, and $n=0,...,N-1$ is the angular slit label. Note that for the simplest scenario $N=2$ slits, we recover the results presented in Kumar \emph{et al.} PRL 2010 \cite{56}. Therefore, there are in principle $N^2$  alternative pathways, represented by the two-photon path diagrams described in  Fig. 1(b), by which the down-converted photons can
pass through the apertures and get detected in coincidence
at single-photon avalanche detectors $D_{s}$ and $D_{i}$. The $N^2$ alternative paths here labelled by the index $q=1,..., N^2$, can be expressed as the tensorial product of the subspaces corresponding to each photon $(s,i)$ passing through the slits  $n=0,...,N-1$, respectively, in the form:
\begin{eqnarray}
& &|s,0\rangle \otimes \{|i,0\rangle, |i,1\rangle,..., |i,N-1\rangle\};\nonumber\\
& &|s,1\rangle \otimes \{|i,0\rangle, |i,1\rangle,..., |i,N-1\rangle\};....\nonumber \\ 
& &|s,N-1\rangle \otimes \{|i,0\rangle, |i,1\rangle,..., |i,N-1\rangle\}. \nonumber\\
\end{eqnarray}

Due to the strong correlation between the position of the two photons in the image
plane of the crystal, only paths of the form $|i,n\rangle|s,n\rangle$ will have a significant contribution \cite{56}. Therefore, the two-photon state is approximately in a pure maximally entangled state of the form $|\psi_{n}\rangle=\sum_{n=0}^{N-1} \sqrt{\rho_{nn}}|i,n\rangle|s,n\rangle$. The density matrix of the
qudit state can be expressed in  the following form:
\begin{equation}
    \hat{\rho}= \sum_{n=0}^{N-1}\sum_{m=0}^{N-1} \rho_{nm} |s,n\rangle |i,n\rangle \langle s,m| \langle i,m|,
\end{equation}

where the subindices $(n=0,..N-1)$ and $(m=0,..,N-1)$ label the angular slits for each photon, and the normalization condition imposes  $\mathrm{Tr}[\hat{\rho}]=\sum_{n=0}^{N-1}\rho_{nn}=1$. The off-diagonal terms $\rho_{nm}$ are complex numbers and can be conveniently expressed as $\rho_{nm}=\sqrt{\rho_{nn}\rho_{mm}}\mu e^{i\theta}$, where $\mu$ is the degree of coherence and $\theta$ is the argument of the coefficient $\rho_{nm}$. Due to Hermiticity of the density matrix, we have  $\rho_{nm}=\rho_{mn}^{*}$.

We can write the density matrix $\hat{\rho}$ in the OAM basis by taking the Fourier transform of the amplitude transmissions for the angular slits  $A_{j,n}$, where ($j=s,i$) is the photon label and ($n=0,..,N-1$) the slit index, as expressed in Eq. (2). For a given path $n$, the two-photon state in the OAM mode basis can be expressed as \cite{56}: 

\begin{eqnarray}
|s,n\rangle |i,n\rangle &=& C\sum_{l}
c_{l}\sum_{l'}\frac{1}{2\pi} \int_{-\pi}^{\pi}d\phi_{s}A_{s,n}(\phi_{s}) e^{-i(l' -l)\phi_{s}}|l'\rangle \nonumber \\
& & \times \sum_{l''}\frac{1}{2\pi} \int_{-\pi}^{\pi}d\phi_{i}A_{i,n}(\phi_{i}) e^{-i(l'' +l)\phi_{i}}|l''\rangle, 
\end{eqnarray}

where $C$ is the normalization factor to ensure $|\psi_{n}|^2=\mathrm{Tr}[\hat{\rho}]=1$. We evaluate $|s,n\rangle|i,n\rangle$ by substituting the expressions for $A_{j,n}(\phi_{j})$. Using the expression for the Fourier transform of the angular amplitude transmission:
\begin{eqnarray}
\tilde{A}_{j,n}&=&\frac{1}{2\pi} \int_{\pi}^{\pi} d\phi A_{j,n}(\phi) e^{-il_{j}\phi}\nonumber \\
\tilde{A}_{j,n}&=& \frac{\alpha e^{-il_{j}\beta n}}{2 \pi} \mathrm{sinc}(\frac{\alpha}{2} l_{j}), \nonumber \\
\end{eqnarray}

where $\mathrm{sinc(x)}=\frac{\sin(x)}{x}$. The coincidence count rate $R_{s,i}$ of detectors $D_{i}$ and $D_{s}$, which consists of the probability that a photon is detected at detector $D_{s}$ in mode $|l_{s} \rangle$, and another photon is detected at detector $D_{i}$ in mode $|l_{i}\rangle$, is given by $R_{si}=\langle l_{i}|_{i} \langle l_{s}|_{s} \hat{\rho} |l_{s}\rangle_{s} |l_{i}\rangle_{i}$. Using Eq.(2), Eq. (4) and Eq. (5) we find:

\begin{eqnarray}
    R_{si}&=&\frac{C^2\alpha^2}{16 \pi^4}     |\sum_{l=-L}^{l=L}c_{l} \mathrm{sinc}((l_{s}-l)\alpha/2) \mathrm{sinc}((l_{i}+l)\alpha/2) |^2 \nonumber \\
    & & \times \sum_{n=0}^{N-1} \sum_{m=0}^{N-1} \rho_{nm} e^{-i\beta(l_{s}+l_{i})(n-m)} \nonumber\\
\end{eqnarray}

For the case of two angular slits ($N=2$), we recover the expression presented in Kumar \emph{et al.} PRL 2010. In our notation, the two-slit basis results in $\{ |s,0\rangle |i,0\rangle, |s,0\rangle |i,1\rangle, |s,1\rangle |i,0\rangle, |s,1\rangle |i,1\rangle \}$. In this case, the coincidence count rate can be written as:

\begin{eqnarray}
    R_{si}&=&\frac{C^2\alpha^2}{16 \pi^4}     |\sum_{l=-L}^{l=L}c_{l} \mathrm{sinc}((l_{s}-l)\alpha/2) \mathrm{sinc}((l_{i}+l)\alpha/2) |^2 \nonumber \\
    & & \times [\rho_{00}+\rho_{11} + 2\sqrt{\rho_{00}\rho_{11}}\mu\cos(\beta(l_{s}+l_{i})+ \theta)],\nonumber\\
\end{eqnarray}

with $\mathrm{Tr}[\rho]=\rho_{00}+\rho_{11}=1$, $\rho_{01}=\sqrt{\rho_{00}\rho_{11}}\mu e^{i\theta} $, and $\rho_{10}=\sqrt{\rho_{00}\rho_{11}}\mu e^{-i\theta} $.\\

The diffraction due to the angular apertures $\alpha$  is described by the diffraction envelopes of the form $|\sum_{l=-L}^{l=L}c_{l} \mathrm{sinc}((l_{s}-l)\alpha/2) \mathrm{sinc}((l_{i}+l)\alpha/2) |^2$. On the  other hand, the multi-path  interference term only depends on the separation between slits $\beta$, and is given by the multiple interference term:

\begin{equation}
\sum_{n=0}^{N-1} \sum_{m=0}^{N-1} \rho_{nm} e^{-i\beta(l_{s}+l_{i})(n-m)}  .
\end{equation}

 By measuring such high order interference fringes, via coincidence detection, we can demonstrate entanglement in a high-dimensional space of dimension $D=N^2$.\\
 
 The visibility $V$ of the interference pattern is quantified by the off-diagonal terms:
 \begin{equation}
     V=2\sqrt{\rho_{nn}\rho_{mm}}\mu
 \end{equation}. 
 
 For a two qubit system ($N=2$), the entanglement can be characterized in terms of the Concurrence \cite{29}, given by $C=\mathrm{max}\{0, \lambda_{1}, \lambda_{2}, \lambda_{3}, \lambda_{4} \}$, where $\lambda_{i}$ $(i=1,2,3,4)$ are the positive eigen-values in descending order of the operator $R$, with $R^2=\sqrt{\rho}\sigma_{y}\otimes \sigma_{y}\rho^{*}\sigma_{y} \otimes \sigma_{y} \sqrt{\rho}$, where $\sigma_{y}$ is a Pauli matrix. For the density matrix in Eq. (4), the Concurrence results equal to the visibility (V) of the angular two-photon interference fringes $C=V=2\sqrt{\rho_{nn}\rho_{mm}}\mu$. 

For the multi-path interference case ($N>2$), the entanglement content can be estimated using Logarithmic Negativity,  via an Entanglement Witness protocol, as discussed in Section 4.

\subsection{Asymmetric slit number ($N \neq M$)}

We now consider the more general case of an asymmetric number of slits $N$ and $M$ for signal and idler, respectively. For perfectly phase-matched down-converted photons, spatial correlations in the plane of the crystal determine that signal and idler can only go through opposite slits, and the state of the two photons is a pure maximally entangled state of the form $|\psi_{n} \rangle=\sum_{n=0}^{N} \sqrt{\rho_{nn}}|s,n\rangle|i,n\rangle$. However, if the photons are not maximally entangled due to imperfect phase matching, signal and idler can go through asymmetric slits $N$ and $M$, and the possible pathways will take the general form $|s,n\rangle|i,m\rangle$.

In the OAM representation, the asymmetric pathways for signal ($s$) and idler ($i$) result in:

\begin{eqnarray}
|s,n\rangle |i,m\rangle &=& C\sum_{l}
c_{l}\sum_{l'}\frac{1}{2\pi} \int_{-\pi}^{\pi}d\phi_{s}A_{s,n}(\phi_{s}) e^{-i(l' -l)\phi_{s}}|l'\rangle \nonumber \\
& & \times \sum_{l''}\frac{1}{2\pi} \int_{-\pi}^{\pi}d\phi_{i}A_{i,m}(\phi_{i}) e^{-i(l'' +l)\phi_{i}}|l''\rangle.
\end{eqnarray}

The two-photon state will be in mixed state of the general form:

\begin{equation}
    \hat{\rho}=\sum_{n,n'=0}^{N-1}\sum_{m,m'=0}^{M-1}\rho_{nm,n'm'}|s,n\rangle|i,m\rangle \langle s,n'| \langle i,m'|,
    \end{equation}
    with normalization condition $\mathrm{Tr}[\hat{\rho}]=1$, and Hermiticity condition $\hat{\rho}=\hat{\rho}^{\dagger}$.\\
    
    Using these equations we can derive an expression for the Coincidence Count Rate for mixed states  (asymmetric case $N \neq M$), of the form:
    
    \begin{eqnarray}
    R_{si}&=&\frac{C^2\alpha^2}{16 \pi^4}     |\sum_{l=-L}^{l=L}c_{l} \mathrm{sinc}((l_{s}-l)\alpha/2) \mathrm{sinc}((l_{i}+l)\alpha/2) |^2 \nonumber \\
    & & \times \sum_{n,n'=0}^{N-1} \sum_{m,m'=0}^{M-1} \rho_{nm} e^{-i\beta l_{s}(n-n')}e^{-i\beta l_{i}(m-m')} \nonumber\\
\end{eqnarray}

Note that for the symmetric case $m=n$ and $n'=m'$ we recover the expression obtained for the case $N=M$ (Eq. (7)).
    
\section{Numerical Results}

\subsection{Density matrix graphical representation}

In this Section we present numerical results for the analytical model developed previously. First, we performed graphical representations of the density matrix operator $\hat{\rho}$ in the reduced pathway basis. We note that, the pathway bases $\{|s,n\rangle|i,n\rangle\}$ for pure maximally entangled states  of the form $|\psi_{n}\rangle=\sum_{n=0}^{N-1}\sqrt{\rho_{nn}}|s,n\rangle|i,n\rangle$ are not complete, as they contains only $N$ elements as opposed to $N^2$. The advantage being that by plotting in this reduced basis we represents only  density matrix elements different from zero.  Therefore density matrices of pure maximally entangled states represented in the pathway basis contain only $N \times N$ elements. More specific,  the diagonal elements satisfy  $\rho_{nn}=\frac{1}{N}$ due to normalization condition, and the off-diagonal elements result in $\rho_{nm}=\frac{V}{N}e^{i \theta}$, where $V$ is the visibility of the interference pattern, and is equal to unity $V=1$ for maximally entangled states. As reported in Ref. \cite{56}, the standard technique to obtain the diagonal elements of the density matrix is via Coincidence Counts. On the other hand, aside from the relative phase $\theta$, off-diagonal elements are obtained from the visibility of the interference patterns (see Ref. \cite{56} and References therein, for further details on standard measurement schemes).

In Figure 2 to Figure 5, we display a graphical representation of density matrices $\hat{\rho}$ for pure maximally entangled states with visibility $V=1$, for different phase parameters $\theta=0$ and $\theta=\pi/4$ in the pathway basis  $\{ |s,n\rangle |i,n\rangle \}$, labeled by the indices $(n,m)$ with ($n=0,..,N-1$) and ($m=0,...,N-1$), for a symmetric configuration of slits of dimensions $N=2,4,5,10$, respectively. $\mathrm{Re}[\hat{\rho}]$ is displayed  on the left column and $\mathrm{Im}[\hat{\rho}]$ is presented on the right column. Figures 2(a) and 2(b) ($N=2$), correspond to diagonal elements  $\rho_{nn}=1/N$ and off-diagonal elements $\rho_{nm}=\frac{V}{N}e^{i \theta}$, with $V=1$ and $\theta=0$, respectively.   Figure 2(c) and 2(d) ($N=2$), correspond diagonal elements  $\rho_{nn}=1/N$ and off-diagonal elements $\rho_{nm}=\frac{V}{N}e^{i \theta}$, with $V=1$ and $\theta=\pi/4$, respectively. Next, Figures 3(a) and 3(b) ($N=4$), correspond to diagonal elements $\rho_{nn}=1/N$ and off-diagonal elements $\rho_{nm}=\frac{V}{N}e^{i \theta}$, with $V=1$
 and $\theta=0$, while Figure 3(c) and 3(d) ($N=4$), correspond to diagonal elements $\rho_{nn}=1/N$ and  off-diagonal elements $\rho_{nm}=\frac{V}{N}e^{i \theta}$, with $V=1$
 and $\theta=\pi/4$. Figures 4(a) and 4(b) ($N=5$), correspond to diagonal elements $\rho_{nn}=1/N$ and  off-diagonal elements $\rho_{nm}=\frac{V}{N}e^{i \theta}$, with $V=1$
 and $\theta=0$, while Figure 4(c) and 4(d) ($N=5$), correspond to diagonal elements $\rho_{nn}=1/N$ and off-diagonal elements $\rho_{nm}=\frac{V}{N}e^{i \theta}$, with $V=1$
 and $\theta=\pi/4$. Finally, Figures 5(a) and 5(b) ($N=10$), correspond to diagonal elements $\rho_{nn}=1/N$ and off-diagonal elements $\rho_{nm}=\frac{V}{N}e^{i \theta}$, with $V=1$
 and $\theta=0$, while Figure 5(c) and 5(d) ($N=10$), correspond to diagonal elements $\rho_{nn}=1/N$ and off-diagonal elements $\rho_{nm}=\frac{V}{N}e^{i \theta}$, with $V=1$
 and $\theta=\pi/4$, respectively.

\begin{figure}[b]
\centering
\includegraphics[width=0.5\textwidth]{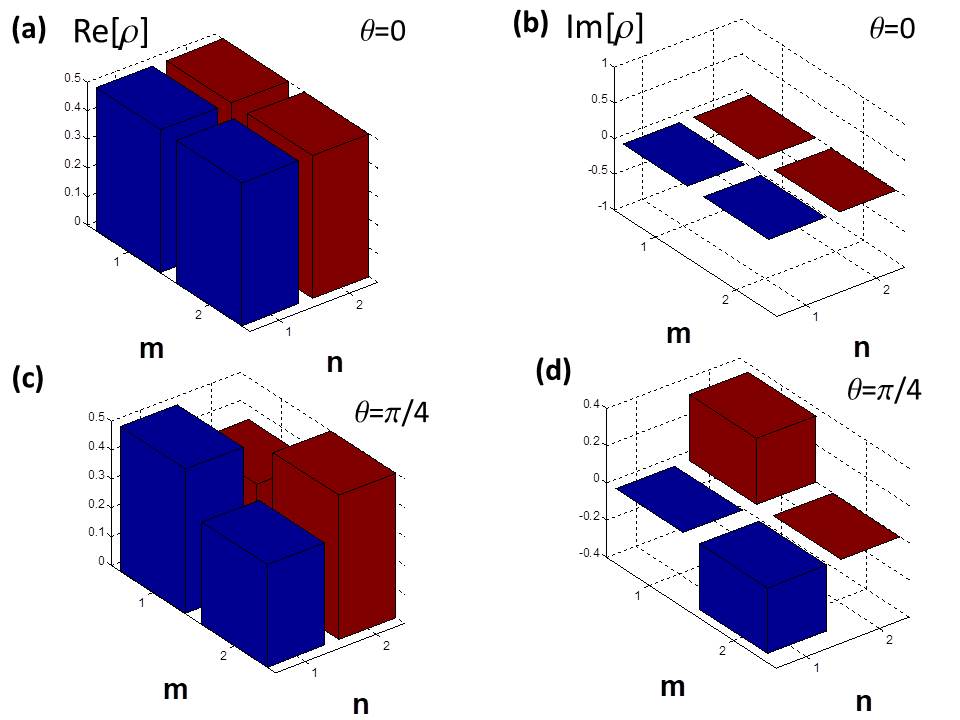}
\caption{Density matrix representation of pure maximally entangled state $|\psi_{n}\rangle$, in the pathway basis $\{|s,n\rangle |i,n \rangle\}$ labelled by indices ($m=0,...,N-1; n=0,...,N-1$), with $N=2$ (see text for details). Left column $\mathrm{Re}[{\hat{\rho}}]$, right column $\mathrm{Im}[{\hat{\rho}}]$, (a) $N=2$, $V=1$, $\theta=0$, (b) $N=2$, $V=1$, $\theta=0$, (c) $N=2$, $V=1$, $\theta=\pi/4$, (d) $N=2$, $V=1$, $\theta=\pi/4$. }
\label{fig:awesome_image}
\end{figure}

\begin{figure}[t]
\centering
\includegraphics[width=0.5\textwidth]{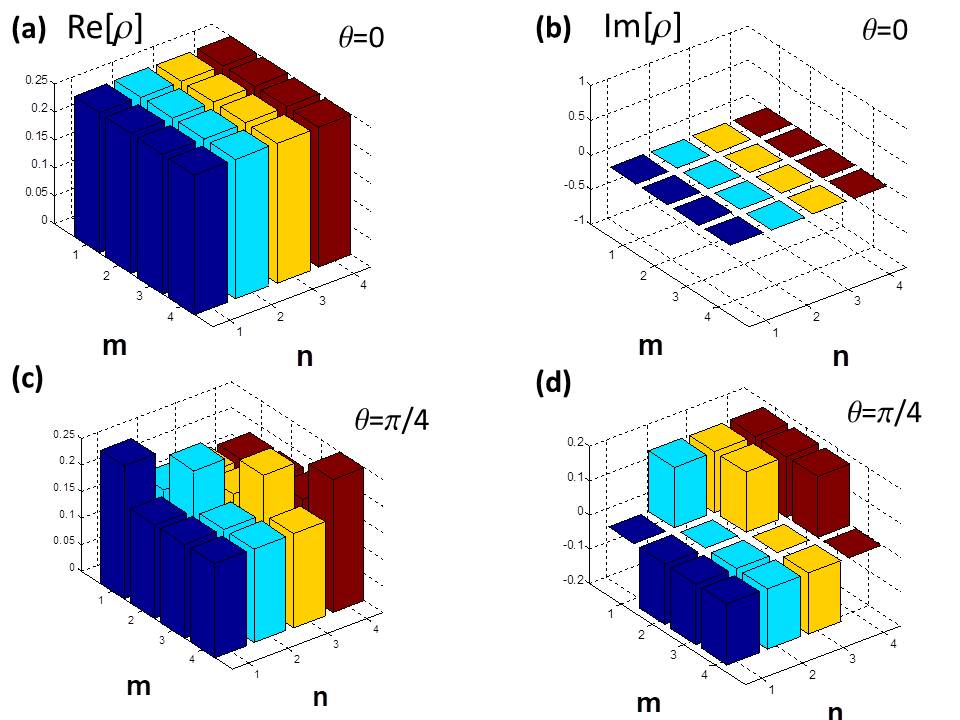}
\caption{Density matrix representation of pure maximally entangled state $|\psi_{n}\rangle$, in the pathway basis $\{|s,n\rangle |i,n \rangle\}$, labelled by indices ($m=0,...,N-1; n=0,...,N-1$), with $N=4$ (see text for details). Left column $\mathrm{Re}[{\hat{\rho}}]$, right column $\mathrm{Im}[{\hat{\rho}}]$, (a) $N=4$, $V=1$, $\theta=0$, (b) $N=4$, $V=1$, $\theta=0$, (c) $N=4$, $V=1$, $\theta=\pi/4$, (d) $N=4$, $V=1$, $\theta=\pi/4$. }
\label{fig:awesome_image}
\end{figure}

\begin{figure}[b]
\centering
\includegraphics[width=0.5\textwidth]{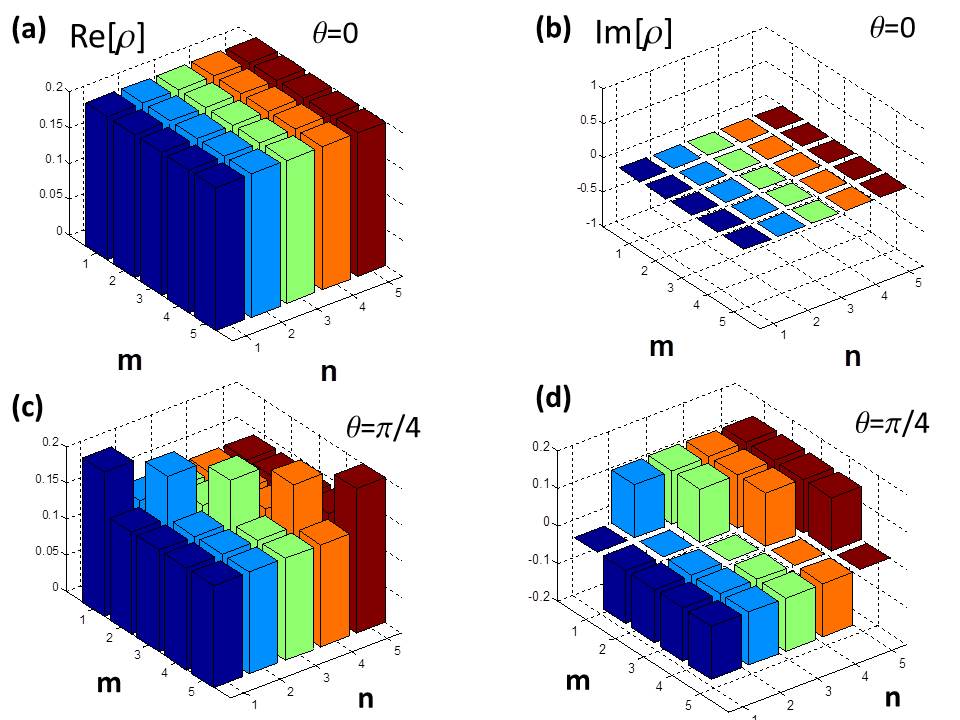}
\caption{Density matrix representation of pure maximally entangled state $|\psi_{n}\rangle$, in the pathway basis $\{|s,n\rangle |i,n \rangle\}$, labelled by indices ($m=0,...,N-1; n=0,...,N-1$), with $N=5$ (see text for details). Left column $\mathrm{Re}[{\hat{\rho}}]$, right column $\mathrm{Im}[{\hat{\rho}}]$, (a) $N=5$, $V=1$, $\theta=0$, (b) $N=5$, $V=1$, $\theta=0$, (c) $N=5$, $V=1$, $\theta=\pi/4$, (d) $N=5$, $V=1$, $\theta=\pi/4$. }
\label{fig:awesome_image}
\end{figure}

\begin{figure}[t]
\centering
\includegraphics[width=0.5\textwidth]{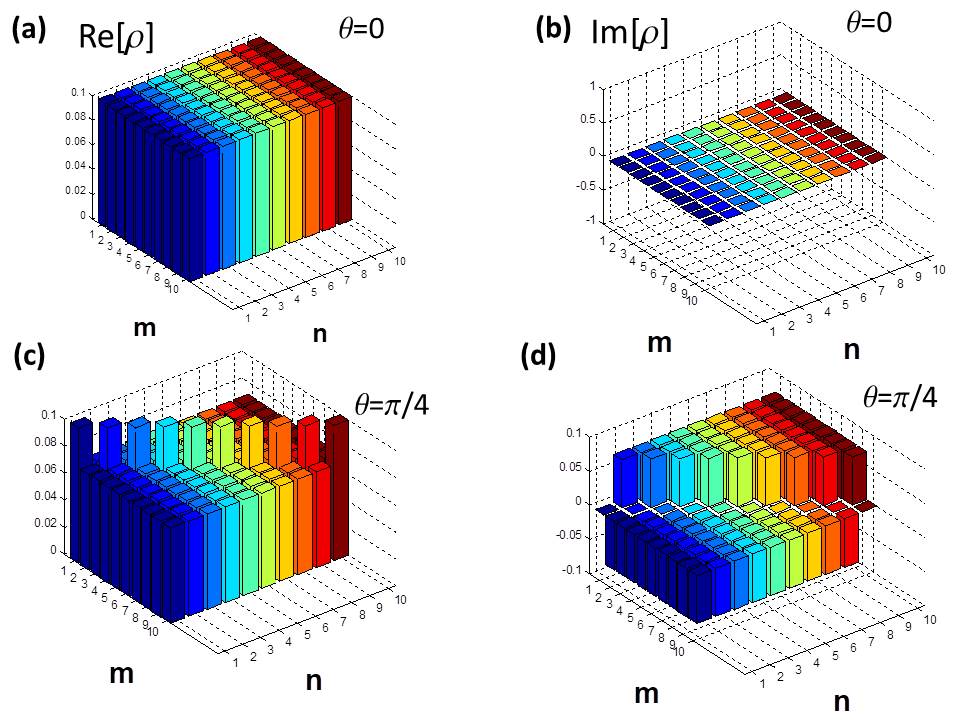}
\caption{Density matrix representation of pure maximally entangled state $|\psi_{n}\rangle$, in the pathway basis $\{|s,n\rangle |i,n \rangle\}$, labelled by indices ($m=0,...,N-1; n=0,...,N-1$), with $N=10$ (see text for details). Left column $\mathrm{Re}[{\hat{\rho}}]$, right column $\mathrm{Im}[{\hat{\rho}}]$, (a) $N=10$, $V=1$, $\theta=0$, (b) $N=10$, $V=1$, $\theta=0$, (c) $N=10$, $V=1$, $\theta=\pi/4$, (d) $N=10$, $V=1$, $\theta=\pi/4$. }
\label{fig:awesome_image}
\end{figure}

\begin{figure}[t]
\centering
\includegraphics[width=0.5\textwidth]{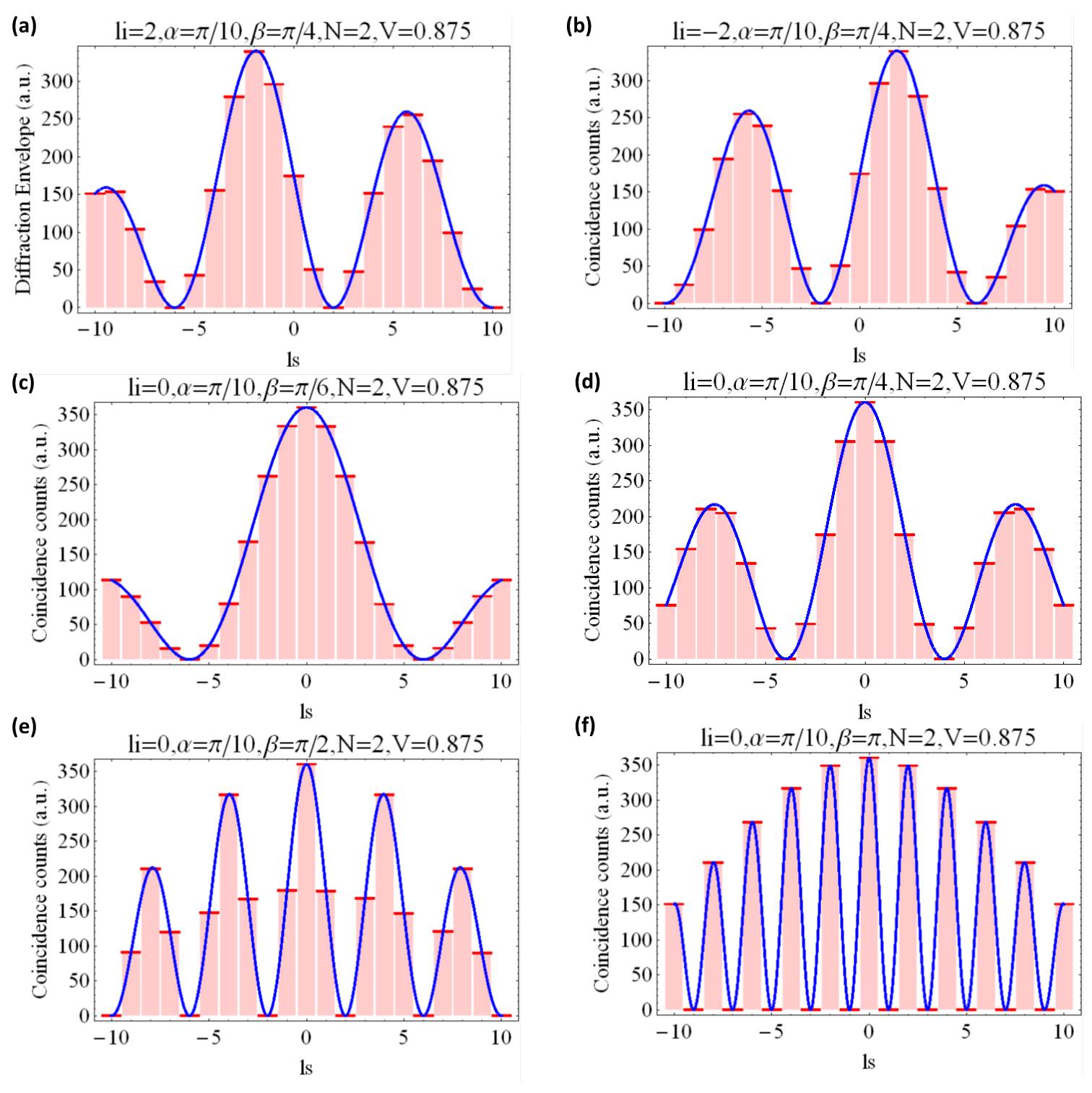}
\caption{Simulated interference fringes, given by  Coincidence Count Rates ($R_{si}$) in Eq. (7), for a reported visibility $V=0.875$, and $N=2$ angular slits. (a) $l_{i}=2$, (b) $l_{i}=-2$. Due to OAM correlations between twin photons the interference pattern has a maximum for $l_{s}=-l_{i}$. Figures  (c)-(f) correspond to different angular separation $\beta$ for $l_{i}=0$ and $\alpha=\pi/10$. (c) $\beta=\pi/6$, (d) $\beta=\pi/4$, (e) $\beta=\pi/2$, (f) $\beta=\pi$. As expected the period of the interference pattern decreases as $\beta$ increases (see Eq.(7) for details). Our numerical results fully reproduce the experimental results reported in Ref [56].}
\label{fig:awesome_image}
\end{figure}

\subsection{Angular Interference for $N=2$ angular slits}

As a starting point, we reproduce the results reported in Kumar \emph{et al.} PRL2010 \cite{56}, for $N=2$ angular slits, resulting in $N^2=4$ alternative pathways. The interference between the alternative paths manifests
itself in the periodic dependence  of the Coincidence Count Rate $R_{si}$, on the angular separation $\beta$ and on the sum of OAMs $l$. We consider $l_{i}=-2,0,2$, $\alpha=\pi/10$,  $\beta=\pi/4, \pi/6, \pi/2, \pi $, and a reported visibility $V=0.875$ [56]. In Figure 6, we present a numerical simulation of Coincidence Count Rate $R_{si}$, given by Eq. (7) for  $\alpha=\pi/10$, $\beta=\pi/4$ and $L=10$. The width of the diffraction envelope increases as the angular aperture $\alpha$ decreases, since angular position and OAM are Fourier related \cite{16, 17}. Therefore the uncertainty in OAM ($\Delta l$) increases as the uncertainty in angular position ($\Delta \phi$) decreases. Figure 6(a) and (b) correspond to $l_{i}=2$ and $l_{s}=-2$, respectively. Due to correlations in OAM of twin photons, the interference pattern is peaked at $l_{s}=-l_{i}$. \\

In Figure 6(c) to 6(f), we present Coincidence Count Rates $R_{si}$ given by Eq. (7), as a function of $l_{s}$ for $l_{i}=0$, and different values of slit separation $\beta$. We consider $N=2$,  $\alpha=\pi/10$, and a reported visibility $V=0.875$. More specific, Fig. 6(c) $\beta=\pi/6$, Fig. 6(d) $\beta=\pi/4$, Fig. 6(e) $\beta=\pi/2$, Fig. 6(f) $\beta=\pi$. As expected the period of the interference pattern decreases as $\beta$ increases (see Eq. (8)). Our numerical results perfectly reproduce the experimental results presented in Kumar \emph{et al.} PRL 2010, which further validates our analytical model.

\subsection{Angular Interference for $N$ angular slits ($N>2$)}

Having verified that our model fully reproduces the experimental results reported in Kumar \emph{et al.} PRL2010, we proceed to the multiple-path interference scenario, for $N>2$. Measurement of such higher order interference fringes, as described by Coincidence Count Rates ($R_{si}$) given by Eq. (7),  can demonstrate path entanglement in high dimensions. We simulated such multi-path interference fringes case, for the cases $N=4, 6, 10$, and $l_{i}=0,2,-2$. We consider $\alpha=\pi/10$, $\beta=\pi/4,\pi/7, \pi/11, \pi/14$  and $V=0.875$. In all cases  the parameters chosen satisfy the condition $N(\alpha+\beta) \leq 2 \pi$. The multi-path interference effect is characterized by  periodic interference patters, where the characteristic period decreases with $\beta$. Due to limited space and visual clarity, we only display results for the case $N=6$.\\

Figure 7 presents numerical simulations of interference fringes as a function of $l_{s}$ ( Eq. (7))  for different values of $l_{i}$, and  slit separation $\beta$, considering $\alpha=\pi/10$, a visibility $V=0.875$, and $N=6$ angular slits, corresponding to a pathway dimension $D=36$. Figure 8(a) $l_{i}=-2$, and figure 8(b) $l_{i}=2$. Figures 7(c)-7(f) display simulated interference fringes, for $l_{i}=0$, $\alpha=\pi/10$, $V=0.875$,  and different angular separations $\beta$ of the form: (c) $\beta=\pi/6$, (d) $\beta=\pi/10$, (e) $\beta=\pi/12$, (f) $\beta=\pi/20$.\\



\begin{figure}[b]
\centering
\includegraphics[width=0.5\textwidth]{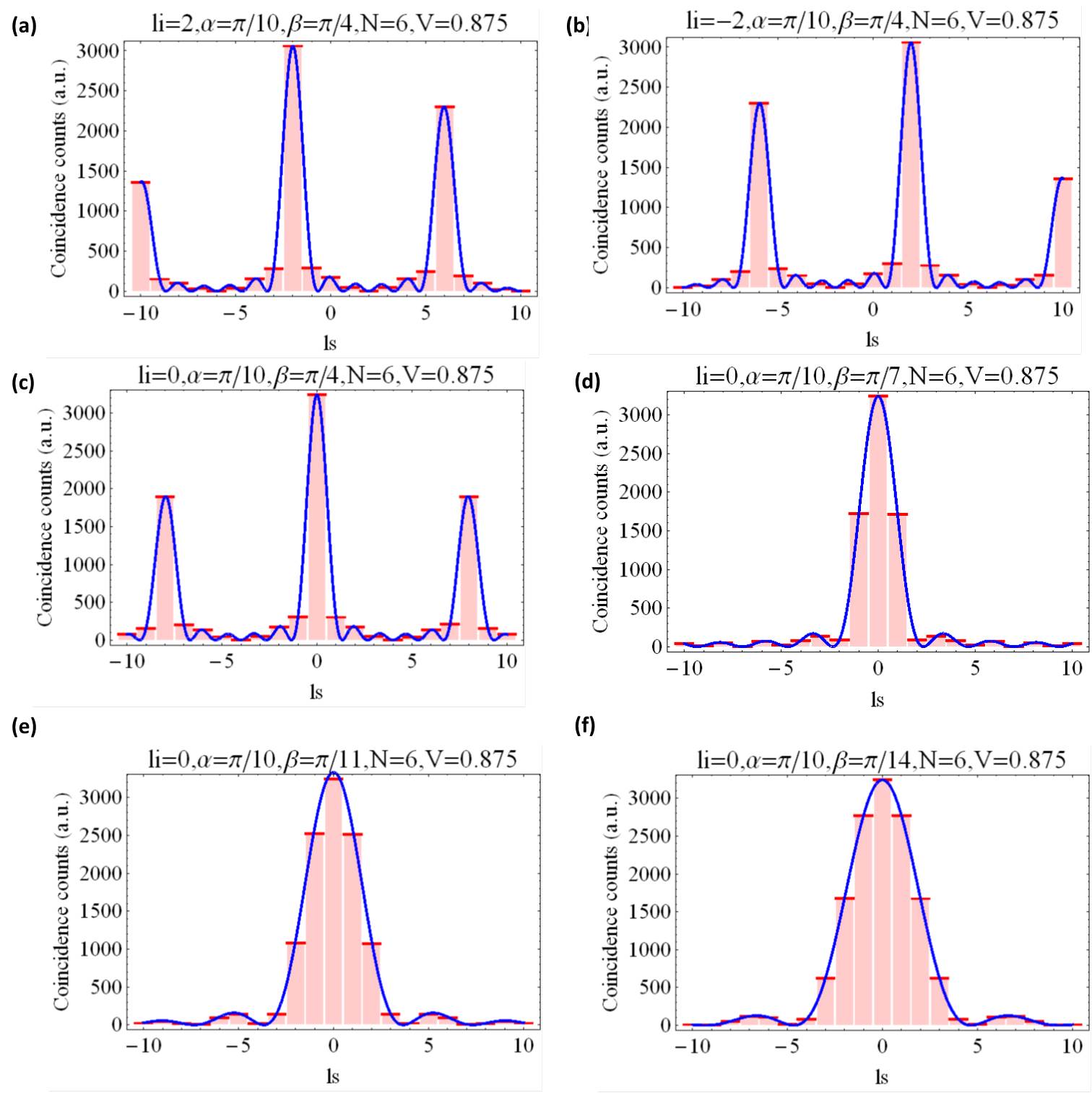}
\caption{Simulated interference fringes, given by  Coincidence Count Rates ($R_{si}$) in Eq. (7), for  visibility $V=0.875$, $\alpha=\pi/10$, $\beta=\pi/4$, and $N=6$ angular slits. (a) $l_{i}=2$, (b) $l_{i}=-2$, Due to OAM correlations between twin photons the interference pattern has a maximum for $l_{s}=-l_{i}$. Figures  (c)-(f) correspond to different angular separation $\beta$, for $l_{i}=0$ and $\alpha=\pi/10$. (c) $\beta=\pi/4$, (d) $\beta=\pi/7$, (e) $\beta=\pi/11$, (f) $\beta=\pi/14$. As expected the period of the interference pattern decreases as $\beta$ increases (see text for details). }
\label{fig:awesome_image}
\end{figure}

Finally, Figure 8 displays interference fringes as a function of $l_{s}$, for $l_{i}=0$, $\alpha=\pi/10$, for the generic case of asymmetric slit number ($N \neq M$), which can produced mixed pathway entangled states (Eq. (11)), with Coincidence Count Rates given by Eq. (13), such mixed states could be implemented via imperfect phase matching, with  coherence characterized by a visibility $V$. Obviously, for a fully incoherent source, the Coincidence Count Rate should be zero. We consider $N=6$ and $M=3$ angular slits, $V=0.875$ and different slit separations $\beta$. Such interference effects are a signature of  mixed path entanglement in a $D$-dimensional space spanned by different path alternatives of dimension $D=N \times M= 18$. Figure 9 (a) $\beta=\pi/4$, Figure 9 (b) $\beta=\pi/7$. As expected the period of the interference pattern decreases as $\beta$ increases (see Eq. (13) for details). For lower visibility, the number of coincidence counts decreases, but the shape of the interference pattern remains the same. 

\begin{figure}[t]
\centering
\includegraphics[width=0.4\textwidth]{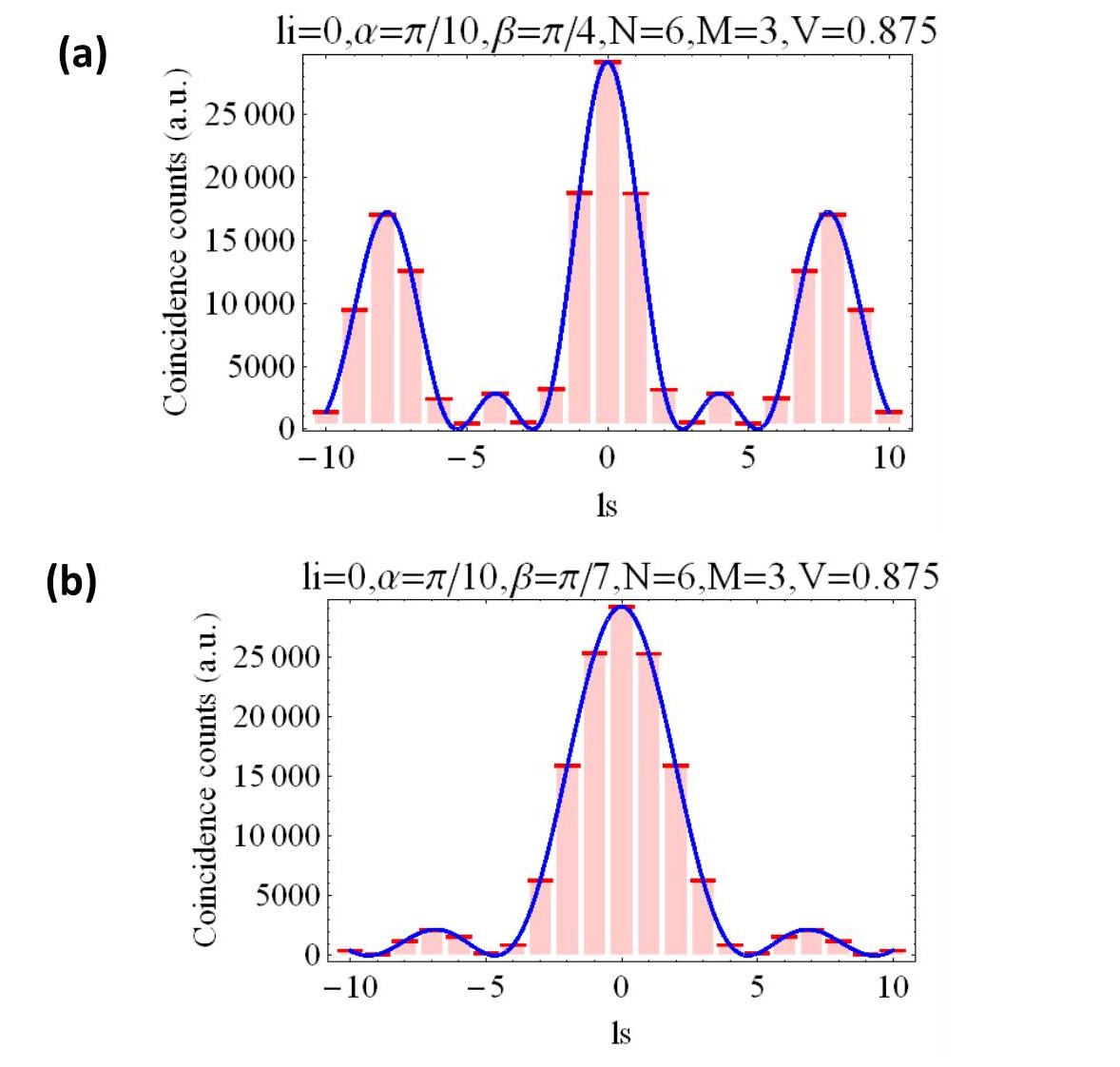}
\caption{Coincidence Count Rate $(R_{s,i})$ given by Eq. (13) for  mixed  states as a function of $l_{s}$, for $l_{i}=0$, $\alpha=\pi/10$, $V=0.875$, $N=6$ and  $M=3$ angular slits (see text for details). We consider different slit separations (a) $\beta=\pi/4$, (b) $\beta=\pi/7$. Such interference effects are a signature of mixed-state path entanglement in a $D$-dimensional space spanned by the different path alternatives of dimension $D=N \times M=18$. As expected the period of the interference pattern decreases as $\beta$ increases (see Eq. 13 for details). }
\end{figure}

\section{Entanglement Witnesses}

For the case of $N=2$ angular slits, the entanglement content can be easily quantified via the Concurrence, in terms of the visibility ($V$) of the interference pattern. For larger spaces ($N>2$), the amount of entanglement can be estimated via an Entanglement Witness. The advantage of the Entanglement Witness approach being that it does not require full tomographic reconstruction of the density matrix, a resource-demanding task, specially for high-dimensional systems. To this end, we are seeking the amount of entanglement in the least entangled physical state compatible with an incomplete set of measurement results. Mathematically, this problem can be presented as

 \begin{equation}
 \label{eq:Emin}
 E_{\min} = \min_{\hat{\rho}}\{E(\hat{\rho}): \tra(\hat{\rho} M_i)=m_i\}, 
 \end{equation}
where $E$ is an entanglement measure, and $M_i$ are the
measurements operators, typically described by a Positive Operator Valued Measurement (POVM),  with measurement data $m_i.$ Additional
constraints are required for $\hat{\rho}$ to be a density matrix, i.e., positive definite, and normalization constraint
$\tra(\hat{\rho})=1$, are also imposed. Depending on the measure of
entanglement, and the measurements $M_i$ chosen, the minimization in Eq.\
(\ref{eq:Emin}) can even be accomplished analytically, generically
that is not the case. Here, we present a protocol based on Refs.\ \cite{ap06,eba07} that allows
this problem to be cast as a semi-definite program when the
entanglement measure is the Logarithmic Negativity~\cite{p05}.\\

Logarithmic Negativity is defined as the logarithm of the 1-norm
of the partial transposed density matrix $\|\hat{\rho}^{T_1}\|_1.$ The
1-norm can be expressed as~\cite{b97}
 \begin{equation}
\|\hat{\rho}^{T_1}\|_1=\max_{\|H\|_{\infty}=1}\tra(H\hat{\rho}^{T_1}) =
\max_{\|H\|_{\infty}=1}\tra(H^{T_1}\hat{\rho}),
 \end{equation}
with the maximization condition over all Hermitian operators $H$, here
$\|.\|_\infty$ denotes the standard matrix operator norm, namely the
largest singular value of the matrix.
Using the monotonicity of the logarithm, the minimization
in Eq.\ (\ref{eq:Emin}) can be rewritten as
  \begin{eqnarray}
 \label{eq:minmax}
 \mathcal{N}_{\min} & = & \log\min_{\hat{\rho}} \{ 
\max_{H}\{\tra(H^{T_1}\hat{\rho})\big| \|H\|_{\infty}=1\}\nonumber \\
& & : \tra(\hat{\rho}
 M_i)=m_i \}.\nonumber \\
  \end{eqnarray}
This  equality allows us to interchange the maximization and
the minimization, leading to
  \begin{eqnarray}
 \label{eq:maxmin}
 \mathcal{N}_{\min} &=& \log\max_{H}\{\min_{\hat{\rho}} 
\{\tra(H^{T_1}\hat{\rho}):\nonumber \\ 
& & \tra(\hat{\rho}
 M_i)=m_i\}: \|H\|_{\infty}=1 \}.\nonumber \\
  \end{eqnarray}
 For any real numbers $\{\nu_i\}$ for which
 \begin{equation}
        H^{T_1}\geq\sum_i\nu_iM_i,
 \end{equation}
 clearly the lower bound on this equation
 \begin{equation}
        \tra(H^{T_1}\hat{\rho})\geq \sum_i\nu_i \tra(M_i\hat{\rho}) = \sum_i\nu_i m_i. 
 \end{equation}
holds true for states $\hat{\rho}$. Thus we get

 \begin{eqnarray}
  \label{eq:maxmax}
\mathcal{N}_{\min} &\geq& \log\max_{H}\{ \nonumber  \\ 
 &  & \times \max_{\nu_i}\big\{\sum_i
\nu_im_i: H^{T_1}\geq\sum_i\nu_iM_i \big\} : \nonumber \\ & & \|H\|_{\infty}=1 \}.\nonumber \\  
  \end{eqnarray} 
  
Note that, at this point, the state $\hat{\rho}$ drops  out of contention now.
Since the inner minimization in Eq.\ (\ref{eq:maxmin}) is a
semidefinite program, strong duality in the feasible case
ensures equality in Eq.\ (\ref{eq:maxmax}). Thus, having fixed the
measurement operators  $M_i,$ any choice of $H$ and
$\nu_i$ such that $H^{T_1}\geq\sum_i\nu_iM_i$ and
$\|H\|_{\infty}=1,$ provides for a lower bound on the
Logarithmic Negativity of states which provide expectation values
of $m_i.$ Finally, we can rewrite Eq.\ (\ref{eq:maxmax}) as
 \begin{eqnarray}
 \label{eq:sdp}
 &&\mathrm{maximize}\;\;\;\log\Big(\sum_i \nu_im_i\Big), \nonumber\\
 &&\mathrm{subject\;to}\;\;H^{T_1}\geq\sum_i\nu_iM_i,\nonumber\\
 && \;\;\;\;\;\; \mathrm{and}\;\;\;   -\mathbb{I} \leq H \leq 
\mathbb{I},\nonumber
  \end{eqnarray}\nonumber
which can be solved relatively easily using standard convex optimization approaches, once the measurement operators
$M_i$ are selected. Since these measurement operators are to be local, the typical form of the
measurement, in the case of bipartite states, such as for signal ($s$ and idler photons), is
  \begin{equation}
 \label{eq:localmeas}
M_n = \Pi^{s}_j\otimes\Pi^{i}_k.
  \end{equation}
 The problem is thus reduced to the construction of the local operators
$\Pi^{s,i}_j$. In
passing, we state that the choice of these measurement operators
can also be cast as a convex optimization problem, although it is more challenging to
incorporate the locality constraint into its framework. For the case a bipartite state given by Eq. (1), it is apparent that the natural set of operators $\Pi^{s,i}_j$ are projectors in the OAM basis, of the form:

\begin{equation}
\Pi^{s,i}_j=|l_{j}\rangle \langle l_{j}|^{s,i}.
\end{equation}

This idea gives useful and practically tight bounds to the
entanglement content, without assuming any prior knowledge
about the state, or its  properties, such as its purity. If the
set of expectation values $\{ \tra(M_n\hat{\rho}) \}$ is tomographically
complete, obviously, the bound gives the exact
value, but in practice, a much smaller number of measurements is
sufficient to arrive at good bounds. Data of expectation values
can be composed, that is if two sets of expectation values are
combined, the resulting bound can only become better, to the
extent that two sets that only give rise to trivial bounds can
provide tight bounds. The approach presented here is 
suitable for any finite-dimensional system,  as long as the observables $M_i$ are
bounded operators.

\section{Experimental Implementation}

The proposed experiment to demonstrate high-dimensional interference and entanglement using $N$ angular slits is depicted in Figure 1(a), and it is based on the experimental setup described in Refs. \cite{18, 56}. In this setup, the pump is a frequency-tripled, mode-locked, Nd-YAG laser  with a pulse repetition frequency of 100 MHz at 355 nm. SLM denotes a
state-of-the-art Spatial Light Modulator, SMF a single
mode fiber, and F an interference filter with 10-nm bandwidth, centered at 710 nm. A 400 $\mu$m diameter Gaussian
pump beam is normally incident on a 3-mm-long crystal
of beta barium borate (BBO), phase matched for frequency degenerate type-I down-conversion with a typical semi-cone angle of the
down-converted beams of 3.5 degrees. For the given
pump beam and phase-matching parameters, the conservation of OAM is strictly obeyed in the down-conversion
process \cite{56}. The main novel ingredient in the setup is given by the angular masks containing $N$ angular slits.\\

Anuglar aperture masks are placed in the path of signal and  idler down-converted  photons, produced by a pump beam with a Gaussian profile with zero OAM ($l=0$), as depicted in Figure 1(c). The generated OAM spectrum transmitted through the angular apertures is analyzed in terms of transmitted spiral harmonics, typically over a range from $l=-12$ to $l=12$. Standard Spatial Light Modulators (SLMs) are used both for preparing the state via the angular apertures, and for analysing the resulting modes \cite{18} (Figure 1(c)). As it is well known in the literature \cite{18}, SLMs are programmable refractive elements, which enable full control of the amplitudes of the diffracted beams. In the standard technique, if the index of the analysis $l$-forked hologram is opposite to that of the incoming mode, planar wave-fronts with on-axis intensity are generated in the first difffraction order. The on-axis intensity can be coupled to single-mode fibers with high efficiency, and can be measured with single-photon detectors $D_{s,i}$, using a coincidence count circuit  (see Figure 1(a) for details). \\

The maximum number of angular slits $N$ - and possible paths $D=N^2$ - that can be implemented in a setup as described in Fig. 1(a), will be determined by the resolution of the Spatial Light Modulators. For a state-of-the-art modulator, with a diameter consisting of  $D=2643$  pixels, and a pixel size $d=3.74 \mu m$, the smallest angular slit aperture ($\alpha$) that can be implemented corresponds to $\alpha/2=\arctan[2d/D]$, which is approximately $\alpha=\pi/2000$. For $\alpha=\beta=\pi/2000$, the maximum number of angular slits $N$ results $N=2\pi/(\alpha+\beta)=2000$, therefore the highest dimension of the qudit space that can be implemented becomes $D=N^2=4000000$, which is several orders of magnitude higher than the largest Hilbert space ever realized with photons, cold atoms, or cold ions. \\

\section{Discussion}

Higher dimensional entangled states are a fundamental resource both from the foundations of quantum mechanics perspective and for the development of new protocols in quantum communication. Maximally entangled states of bipartite quantum systems in an N-dimensional Hilbert space, the so called qudits,  can introduce higher violations of  local realism than qubits \cite{44}, and can prove more resilent to noise than qubits \cite{44,45}. In quantum cryptography \cite{46}, or other quantum information protocols \cite{63,57,58,59,60,61}, use of  entangled qutrits ($N = 3$) \cite{47, 48} or qudits \cite{49,50} instead of qubits is more secure against attacks. Moreover, it is known that quantum protocols work best for maximally entangled states. These facts motivate the development of techniques to generate maximally entangled states in higher dimensional Hilbert spaces. Entangled qutrits with two photons using an unbalanced 3-arm fiber optic interferometer \cite{53}  has been demonstrated. Time-bin entangled qudits up to $D = 11$ from pump pulses generated by a mode-locked laser has also been reported \cite{55}. Here we report a protocol that can produce entangled qudits, based on angular diffraction, with a maximal dimension $D=4000000$, only limited by the resolution of the Spatial Light Modulators. 

\bigskip

\section{Acknowledgements}
The author is grateful to Sonja Franke-Arnold and Leonardo Neves for useful discussions.
GP acknowledges  J. Eisert for valuable insights into Entanglement Witnesses and  convex optimization protocols. GP acknowledges financial supports via grants PICT Startup 2015 0710, and UBACyT PDE 2017.


\begin{thebibliography}{4}
\bibitem{Griffiths}
D. J. Griffiths,
\textit{Introduction to Electrodynamics}
(Cambridge University Press, Cambridge, 2017).
\bibitem{1}C. K. Hong, Z. Y. Ou, and L. Mandel, Phys. Rev. Lett. \textbf{59},
2044 (1987).
\bibitem{3} T. J. Herzog et al., Phys. Rev. Lett. \textbf{72}, 629 (1994).
\bibitem{2} A. K. Jha et al., Phys. Rev. A \textbf{77}, 021801(R) (2008).
\bibitem{5} J. Brendel et al., Phys. Rev. Lett. \textbf{82}, 2594 (1999).
\bibitem{6} R. T. Thew et al., Phys. Rev. A \textbf{66}, 062304 (2002). 
\bibitem{4} E. J. S. Fonseca et al., Phys. Rev. A \textbf{61}, 023801 (2000).

\bibitem{8}L. Neves et al., Phys. Rev. Lett. \textbf{94}, 100501 (2005).
\bibitem{7} L. Neves et al., Phys. Rev. A \textbf{76}, 032314 (2007).
\bibitem{10} A. Aspect, P. Grangier, and G. Roger, Phys. Rev. Lett. \textbf{49},
91 (1982).
\bibitem{9} L. Mandel, Rev. Mod. Phys. \textbf{71}, S274 (1999).
\bibitem{12} A. Zeilinger, Rev. Mod. Phys. \textbf{71}, S288 (1999).
\bibitem{11} A. K. Ekert, Phys. Rev. Lett. \textbf{67}, 661 (1991).
\bibitem{14} C. H. Bennett and S. J. Wiesner, Phys. Rev. Lett. \textbf{69}, 2881
(1992).
\bibitem{15} C. H. Bennett et al., Phys. Rev. Lett. \textbf{70}, 1895 (1993).
\bibitem{13} S. M. Barnett and D. T. Pegg, Phys. Rev. A \textbf{41}, 3427
(1990).
\bibitem{17} S. Franke-Arnold et al., New J. Phys. \textbf{6}, 103 (2004).
\bibitem{16} B. Jack, M. Padgett, and S. Franke-Arnold, New J. Phys.
\textbf{10}, 103013 (2008).
\bibitem{19} A. K. Jha et al., Phys. Rev. A \textbf{78}, 043810 (2008).
\bibitem{18} A. Vaziri, G. Weihs, and A. Zeilinger, Phys. Rev. Lett. \textbf{89}, 240401 (2002).
\bibitem{22} N. K. Langford et al., Phys. Rev. Lett. \textbf{93}, 053601 (2004).
\bibitem{20} J. Leach et al., Opt. Express \textbf{17}, 8287 (2009).
\bibitem{21} P. G. Kwiat et al., Phys. Rev. Lett. \textbf{75}, 4337 (1995).
\bibitem{25} S. Ramelow et al., Phys. Rev. Lett. \textbf{103}, 253601 (2009).
\bibitem{24} J. G. Rarity and P. R. Tapster, Phys. Rev. Lett. \textbf{64}, 2495
(1990).
\bibitem{23} M. N. O’Sullivan-Hale et al., Phys. Rev. Lett. \textbf{94}, 220501
(2005).
\bibitem{27} S. P. Walborn et al., Phys. Rev. A \textbf{69}, 023811 (2004).
\bibitem{26} S. Franke-Arnold et al., Phys. Rev. A \textbf{65}, 033823 (2002).
\bibitem{28} J. P. Torres, A. Alexandrescu, and L. Torner, Phys. Rev. A
\textbf{68}, 050301(R) (2003).
\bibitem{30} W. K. Wootters, Phys. Rev. Lett. \textbf{80}, 2245 (1998).
\bibitem{29} B. Jack et al., New J. Phys. \textbf{11}, 103024 (2009).
\bibitem{31} G. Molina-Terriza, J. Torres, and L. Torner, Opt. Commun.
\textbf{228}, 155 (2003).
\bibitem{34} A. Mair et al., Nature (London) \textbf{412}, 313 (2001).
\bibitem{33} J. Leach et al., New J. Phys. \textbf{7}, 55 (2005).
\bibitem{32} G. Tyler and R. Boyd, Opt. Lett. \textbf{34}, 142 (2009).

\bibitem{ap06}
        K.\ M.\ R.\ Audenaert and M.\ B.\ Plenio, New J.\ Phys.\ \textbf{8}, 266 (2006).

\bibitem{eba07}
        J.\ Eisert, F.\ G.\ S.\ L.\ Brand{\~a}o, and K.\ M.\ R.\ Audenaert,
        New J.\ Phys.\ \textbf{8}, 46 (2007).

\bibitem{grw07}
        O.\ G\"{u}hne, M.\ Reimpell, and R.\ F.\ Werner, Phys.\ Rev.\ Lett.\ {\bf 98}, 110502 (2007).

\bibitem{p09}
        M.\ B.\ Plenio, Science, \textbf{324}, 342 (2009).

\bibitem{p05}
        M.\ B.\ Plenio, Phys.\ Rev.\ Lett.\ \textbf{95}, 090503 (2005);
        J.\ Eisert, PhD thesis (Potsdam, February 2001);
         G.\ Vidal and R.F.\ Werner,
         Phys.\ Rev.\ A {\bf 65}, 032314 (2002).

\bibitem{Eisert}  D. E. Browne, J. Eisert, S. Scheel, and M. B. Plenio,
        Phys. Rev. A \textbf{67}, 062320 (2003);
        J. Eisert, D.~E. Browne, S. Scheel, and M.~B. Plenio,
        Annals of Physics (NY) {\bf 311}, 431 (2004).

\bibitem{distillation}
        J. Eisert, S. Scheel, and M.~B. Plenio, Phys. Rev. Lett. {\bf 89}, 137903 (2002);
        J. Fiurasek, Phys. Rev. Lett. {\bf 89}, 137904 (2002);
        G. Giedke and J.~I. Cirac, Phys. Rev. A {\bf 66}, 032316 (2002).

\bibitem{b97}
        R. Bhatia, \emph{Matrix analysis}, Springer, New York (1997).


\bibitem{45} D. Kaslikowski et al., Phys. Rev. Lett. \textbf{85}, 4418 (2000).
\bibitem{44} D. Collins et al., Phys. Rev. Lett. \textbf{88}, 040404 (2002).
\bibitem{47} A. K. Ekert, Phys. Rev. Lett. \textbf{67}, 661, (1991).
\bibitem{46} H. Bechmann-Pasquinucci and A. Peres, Phys. Rev. Lett.
\textbf{85}, 3313 (2000).
\bibitem{49} T. Durt, N. J. Cerf, N. Gisin, and M. Zukowski, Phys.
Rev. A \textbf{67}, 012311 (2003).
\bibitem{48} M. Bourennane, A. Karlsson, and G. Bj¨ork, Phys. Rev.
A \textbf{64}, 012306 (2001).
\bibitem{50} N. J. Cerf, M. Bourennane, A. Karlsson, and N. Gisin,
Phys. Rev. Lett. \textbf{88}, 127902 (2002).
\bibitem{52} C. H. Bennett et al., Phys. Rev. Lett. \textbf{70}, 1895 (1993).
\bibitem{51} J. C. Howell, A. Lamas-Linares, and D. Bouwmeester,
Phys. Rev. Lett. \textbf{85}, 030401 (2002).
\bibitem{53} R. T. Thew, A. Ac´ın, H. Zbinden and N. Gisin, Phys.
Rev. Lett. \textbf{93}, 010503 (2004).
\bibitem{55} A. Vaziri, G. Weihs, and A. Zeilinger, Phys. Rev. Lett.
\textbf{89}, 240401 (2002).
\bibitem{54} H. de Riedmatten, I. Marcikic, H. Zbinden and N. Gisin,
Quant. Inf. and Comp. \textbf{2}, 425 (2002).
\bibitem{56} A. Kumar Jha, J. Leach, B. Jack, S. Franke-Arnold, S. Barnett, R. Boyd, M. Padgett,
Phys. Rev. Lett.  \textbf{104}, 010501 (2010).

\bibitem{63}G. Puentes, A. Datta, A. Feito, J. Eisert, M.B. Plenio, I.A. Walmsley, New Journal of Physics \textbf{12}, 033042 (2010).

\bibitem{57} G. Puentes, G. Waldherr, P. Neumann, G. Balasubramanian, J. Wrachtrup, 
Scientific Reports \textbf{4}, 1-6 (2014).

\bibitem{58} G. Puentes, A. Aiello, D. Voigt, J.P. Woerdman, 
Physical Review A \textbf{75}, 032319 (2007).

\bibitem{61} G. Puentes, G. Colangelo, R.J. Sewell, M.W. Mitchell, New Journal of Physics \textbf{15}, 103031 (2013).

\bibitem{59} O. Takayama, J. Sukham, R. Malureanu, A.V. Lavrinenko, G. Puentes, Optics letters \textbf{43}, 4602-4605 (2018).

\bibitem{60} S. Moulieras, M. Lewenstein, G. Puentes, Journal of Physics B: Atomic, Molecular and Optical Physics \textbf{46}, 104005	(2013).

\bibitem{62} The numerical codes used for the simulations in this work were programmed in Mathematica, and they are available upon request.

\end{thebibliography}
\end{document}